# Precise Monte Carlo Simulation of Single-Photon Detectors

Mario Stipčević[1,2,*] and Daniel J. Gauthier[1]
[1]Duke University, Department of Physics, Box 90305, Durham, North Carolina 27708, USA
[2]On leave from: Rudjer Boskovic Institute, Bijenicka 54, 10002 Zagreb, Croatia
*corresponding author e-mail: Mario.Stipcevic@phy.duke.edu

*Abstract*—We demonstrate the importance and utility of Monte Carlo simulation of single-photon detectors. Devising an optimal simulation is strongly influenced by the particular application because of the complexity of modern, avalanche-diode-based single-photon detectors.. Using a simple yet very demanding example of random number generation via detection of Poissonian photons exiting a beam splitter, we present a Monte Carlo simulation that faithfully reproduces the serial autocorrelation of random bits as a function of detection frequency over four orders of magnitude of the incident photon flux. We conjecture that this simulation approach can be easily modified for use in many other applications.

## 1. Introduction

Sensitive detectors capable of detecting a single photon, so called "single-photon detectors" or "single-photon counters," are widely used in the laboratory research environment and medical equipment, such as in LIDAR, particle sizers, blood testers, time resolved spectroscopy, nuclear and particle physics and quantum information research. Photon detection usually starts with a conversion of the photon into a single charge carrier in a suitable medium and subsequent charge multiplication. The charge multiplication process results in a sizeable charge or current, which allows for further processing.

Ideally, a photon detector produces one logical pulse, of standard height and width, per each photon received and in perfect synchronization with arrival of the photon. However, in reality, due to the limitations related to the charge multiplication process and imperfections of the electronic circuits required to amplify and shape the signal, there are many deviations from this ideal picture. The most important imperfections include: 1) after each detection (of a photon), the detector is not capable of detecting the next photon(s) received during a certain time interval called the "dead time" or "photon pair resolution time;" 2) after each detection, there is a certain probability of an "afterpulse," an echo detection caused solely by previous detections and not by a real photon; 3) "twilight events" - a photon or afterpulse detection that occurs during the last part of the dead time; 4) in total darkness, the detector produces false detections called "dark counts," which occur randomly in time; 5) the time between the photon arrival time and the electrical pulse has fluctuations (timing jitter); and 6) the detection efficiency is less than unity and is a function of wavelength. Devising an optimal simulation is strongly influenced by the particular application because of the complex operation of photon detectors. In this work, we present simulation of a seemingly very simple application: generation of random numbers by means of a beam splitter, which is important for applications in quantum communication [6-8] and true random number generation [9-11]. But, surprisingly, as we will show, this particular application is very sensitive to most imperfections of single-photon detectors and its successful modeling (simulation) presents quite a challenge.

## 2. Single-photon detector

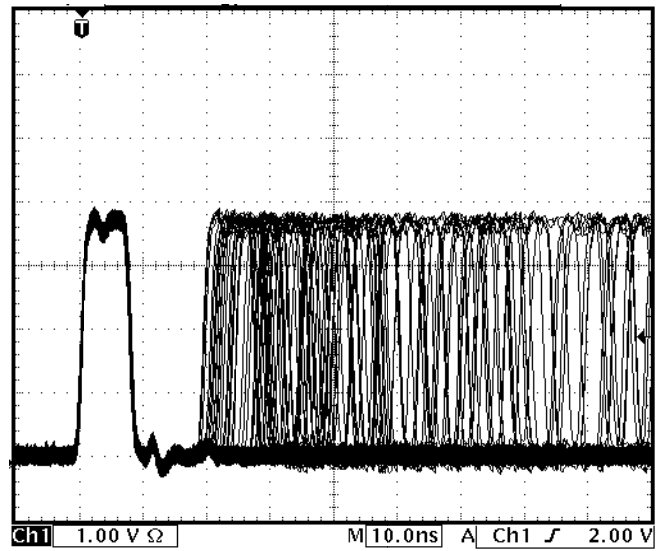

**Figure 1.** Cumulative output of a single-photon detector illuminated by Poissonian light from a LED diode attenuated to yield approximately 250,000 counts per second on average. The dead time is defined a minimum time delay between the trigger pulse (on the right) and the next pulse, and is about 20ns.

In this work, we use two single-photon detectors based on the Excelitas (formerly Perkin Elmer) SLiK silicon avalanche photodiode matched with a custom-built active quenching circuit to form a fully operational photon counting module. The APDs are operated at 19.5V beyond the Geiger threshold at a temperature of -10°C. The dead time of our electronics is about 20ns. Figure 1 shows a typical cumulative output from a single-photon detector illuminated by a Poissonian-distributed stream of photons which appear randomly in time.

## 3. Experimental setup

The experimental setup is depicted in Fig. 2. A red LED diode with an adjustable 3-lens collimator (Roithner RC-LED-650-02 module with Hamamatsu LED L7868, wavelength 670nm), is operated at a constant current of 2.0 mA), coupled to a single-mode fiber, and incoupled to a 50-50 fiber beam splitter (Thorlabs FC632-50B-FC) via kinematic adjustable attenuator. Each arm of the beam splitter is coupled to a single photon detector, denoted as D0 or D1. The arm leading to the detector D0 is attenuated by a factor of 10, while the other arm is attenuated by a variable attenuator, which enables adjusting the ratio of the counting rates of the two detectors to unity. The ratio was directly monitored by a frequency-ratio counter (Hameg, HM8132) and was adjusted to (1.00 ± 0.01).

Ideally, a photon entering the beam splitter randomly decides to be directed to either D0 (producing a bit value "0") or D1 (producing a bit value "1"). In order to obtain random numbers (binary bits), we use a bit-resolving logic circuit (Fig. 3), which generates logical 1 (HIGH) if D1 fires alone, 0 if D0 fires alone, and nothing if pulses from D1 and D0 overlap over a coincidence window of about 12ns. Rejecting coincident events is a precaution to avoid bias that may result from path delays and threshold differences in the bit-resolving circuit even though experimentally we did not find any noticeable difference (*i.e.*, we see no difference when removing the last AND gate before *Strobe*). The resolver has two outputs: *Bit*, which takes on value 0 when D0 fires or value 1 if D1 fires, and *Strobe*, which creates a 3ns wide logic pulse whenever there is new random data waiting to be read. The resolver is implemented in fast Programmable Logic Device (PLD, Altera EPM7128BTC100-4) integrated with an USB2 controller for transferring data to a PC.

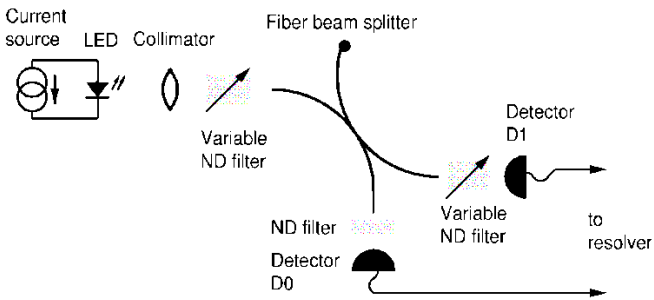

**Figure 2.** Experimental setup of the beam-splitter based optical random number generator.

In order to measure and visualize the response of a detector, we use a time-to-amplitude converter [1], which measures time intervals between two adjacent detections of a single detector and sends the result to a PC. A histogram of time intervals obtained from detector D0 operating at 20kHz detection rate is shown in Fig. 4. The flat background on the right hand side of the picture is caused by detection of real photons. Because the LED is operated in a constant-current mode and a very tiny fraction of photons is coupled into the detector, the photon statistics obeys an exponential probability density function.

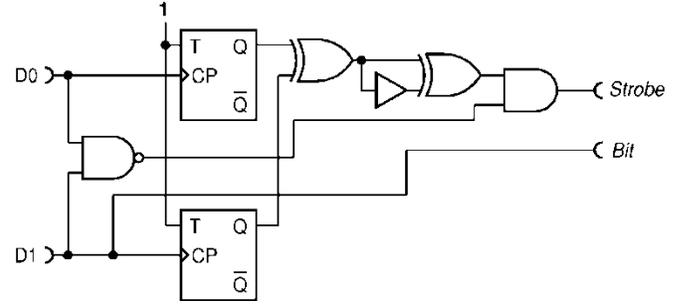

**Figure 3**. Random bit-resolving electronics implemented in the Programmable Logic Device made in 2.5V CMOS technology with input voltage levels compatible with outputs of the detectors.

A computer program is used to subtract this "background," leaving only pathological detections, which consists of afterpulses and twilight events.

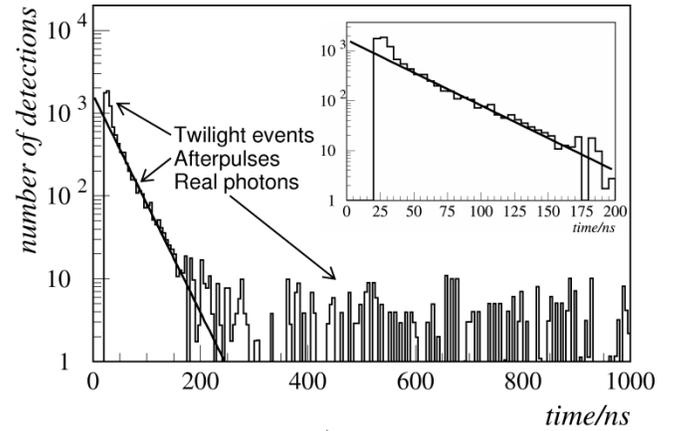

**Figure 4.** Histogram of pulse pair time intervals recorded from detector D0 (background subtracted). Three types of events are clearly visible: detection of real photons, afterpulses and twilight events (see the text). Also shown is exponential fit of afterpulses. The shortest time interval between two subsequent pulses is equal to the dead time (20 ns) as clearly visible in the inset that shows zoom around zero.

In an APD that features only a direct trap decay mechanism, or several independent direct decay mechanisms, one expects an exponential probability density function for the afterpulses [3], which is indeed the case with the SLiK diode used here. This is not generally the case: some APDs have more complex afterpulsing mechanisms, but, by measuring them precisely, one can easily model any afterpulse distribution. Detector D0 shows an exponential decay of afterpulses (linear fit in the log plot) with lifetime of $\tau_a(D0)=33$ ns and afterpulsing probability $P_a(D0)$ of 0.047. Each APD is slightly different: Detector D1 shows a similar behavior as D0 but with afterpulsing lifetime $\tau_a(D1)=40$ ns and $P_a(D1)=0.043$.

Furthermore, on top of afterpulses, we see that there is an excess of events happening right after the dead time. These are the so-called "twilight" events. Namely, the APD quenching pulse starts right at the beginning of the dead time and has a flat-top that lasts for about 12 ns. After that the APD operating voltage is restored to the nominal value during next 4-5 nanoseconds. The avalanche sensing electronics is intentionally shut off from the beginning of the quenching all the way to a few nanoseconds after the voltage on the APD has reached its nominal value in order to avoid oscillations. This whole period of time plus the propagation delay time through the sensing electronics is called the "dead time" because there is no possibility of an output pulse during all that time. The fraction of the dead time during which APD is biased above Geiger breakdown and, therefore, is at least partially sensitive to real photons and afterpulses is called the twilight region. A twilight event eventually produces an output pulse but it appears shifted to *after* the dead time. Because the APD may go into an avalanche sooner or later during the twilight and the recovery time of the electronics may be affected by it, twilight events may appear a bit smeared in time (a few nanoseconds typically), as seems to be the case in our detectors.

**4. Measurements and simulation**

The measurements are done by adjusting the light levels reaching detectors D0 and D1 (set so that they produce approximately the same mean pulse rate at the outputs of the detectors) and collecting the stream of random numbers (Fig. 2). We generate random numbers at multiple pulse rates, ranging from 1 kHz to 10 MHz. The two detectors produce pulses that are independently of each other. Namely even though APDs produce a tiny flash of light every time that they avalanche [13] and even if we assume perfect coupling of these photons (~40 photons/sr for Perkin Elmer C30902) into the fiber, the directivity isolation of the used beam splitter is over 55dB ensuring that this type of correlation among detectors is completely negligible in our study. Some pulses belong to detection of real photons, some to afterpulses, and some to twilight events. I the dark counts can be neglected because they are small and indistinguishable from real photon detection events. The bit-resolving electronics, of course, does not distinguish the origin of a particular pulse, so it treats all pulses the same, which leads to errors in randomness, most notably by creating correlations among the bits. In order to understand how independently operating detectors hit by uncorrelated photons can produce correlated bits we need to study the detection process on some detail.

Let us imagine a photon entering the beam splitter and propagating toward detector D0, thus producing bit value "0." If the next photon arrives a long time later (that is, if photon rate is low) there will be a probability (equal to the afterpulsing probability) that D0 generates *another* bit 0 by means of an afterpulse. The same is true for detector D1. Thus, in the limit of low detection rate where the dead time can be neglected, afterpulsing creates a positive correlation in the bit stream whose order-of-magnitude size is approximately equal to the afterpulse probability.

Now let us imagine that the detectors have no afterpulsing, but do have a dead time and that the photon rate is high. Again, consider a photon entering D0. The next photon, of course, has equal probability to end up in either detector, but it has a larger chance of being detected if it goes to D1 because D0 is dead for some time after detection of the previous photon, and so anti-correlation appears. As the photon rate increases, the probability that the "other" detector will fire next (that is, the probability of getting "1" after "0" and "0" after "1") increases, and so does the anti-correlation. As we will see below, both effects are is indeed confirmed by measurements.

In summary, we have two effects in photon-counting detectors that produce correlations among random bits: afterpulsing, which generates positive correlation and is dominant at low detection rates, and dead time, which generates negative correlation and dominates at high detection rates. Naively, there should be a photon rate at which the two effects exactly cancel. This is indeed true for the lowest-order autocorrelation coefficient, but not for the higher-order coefficients. The problem of inevitable correlations with this method of generating random numbers led researchers to quest for methods more resilient to detector imperfections, such as use of periodic pulses of light [9] or photon arrival time information [14-15], but even there detectors cause errors.

In order to simulate the photon detection process, we simply translate the processes given above into a computer language. A subroutine that generates one detection time interval is given below (in BASIC).

```
SUB Detector(f, ap, atau, deadt, dt)
 stau = 1/f
 IF RND <= ap THEN
  dt = -atau*LOG(RND): REM afterpuls
 END IF
 IF dt < deadt THEN
  WHILE tlast < deadt
   dt = -stau*LOG(RND):REM photo-electron
   tlast = tlast + dt
  WEND
  dt = tlast
 END IF
END SUB
```

Here, the function RND returns a uniform real random number in the interval (0,1), f is the mean frequency of detection of real photons; ap is afterpulsing probability; atau is the exponential lifetime of the afterpulses; deadt is the dead time and dt is the output value of the time interval to the next pulse generated by the detector. Function LOG(x) returns natural logarithm of the argument. Note that the order in which afterpulses and real photon pulses are generated is important. First, a time interval for an afterpuls is generated with the afterpulsing probability. If the interval is greater than the dead time, the function returns the time interval. If the afterpuls is not generated or if the time interval is shorter than the dead time, the routine continues and generates one or more real photons until the first photon that survives dead time, and then returns the cumulative time interval.

By direct measurement, we have already seen that our two detectors have slightly different parameters (dead time, afterpulsing probability and lifetime). In our simulation too we can put two different sets of parameters: one for each detector. In real measurements, the detectors operate simultaneously and therefore, in our simulation program, there is an algorithm that determines the order in which the detectors fire, which is a simulation of the bit resolver. The resolved bits are then written to an output file in binary format. The autocorrelation coefficient is calculated by a separate program.

Measurements and simulations of the serial autocorrelation coefficient *a* (defined, for example in Ref. [5]) are summarized in Table 1. The frequency *f* is the mean pulse frequency at the output of each detector.

| *f*/Hz | *a* (meas.) | *a* (sim. 1) | *a* (sim. 2) | *a* (sim. 3) |
|---|---|---|---|---|
| 1k | 0.03047(34) | 0.02470 | 0.02933 | 0.02926 |
| 3k | 0.03013(24) | 0.02467 | 0.02923 | 0.02917 |
| 10k | 0.02978(16) | 0.02445 | 0.02927 | 0.02915 |
| 20k | 0.02940(16) | 0.02411 | 0.02882 | 0.02877 |
| 50k | 0.02847(16) | 0.02336 | 0.02846 | 0.02843 |
| 100k | 0.02708(16) | 0.02222 | 0.02760 | 0.02750 |
| 200k | 0.02429(16) | 0.02001 | 0.02579 | 0.02555 |
| 500k | 0.01914(16) | 0.01294 | 0.02045 | 0.02051 |
| 1M | 0.01088(16) | 0.00176 | 0.01163 | 0.01169 |
| 2M | -0.00496(08) | -0.02055 | -0.00505 | -0.00499 |
| 3M | -0.02061(11) | -0.04135 | -0.02090 | -0.02100 |
| 5M | -0.05156(16) | -0.08001 | -0.05057 | -0.05055 |
| 7.5M | -0.09080(16) | -0.12365 | -0.08476 | -0.09157 |
| 10M | -0.13385(16) | -0.16296 | -0.11566 | -0.13229 |

**Table 1.** Serial autocorrelation coefficient of random bits: measurements and simulations 1-3 as explained in the text. One standard deviation statistical errors on measured values are given in the table while the error on all simulated values is 1.1E-4.

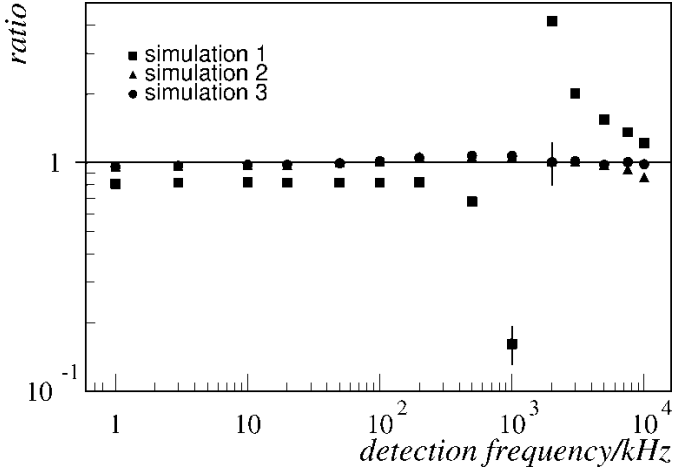

**Figure 5.** Ratios of simulated and measured values of autocorrelation. For each point a ±3 standard deviations error bar is included in the graph.

The most simple simulation (Table 1, simulation 1) is based on the subroutine described above. We see that the general trend of the correlation is followed and change of sign is reproduced correctly. However, simulated values deviate at least 20% from measurements and much more near the point where the two opposing effects (afterpulsing and dead time) strongly interfere causing change of the sign of auto correlation. Too fast a rise of negative correlation at high frequency end and too small positive correlation on the low frequency end can be explained by too long a dead time used in simulation (20 ns).

Apart from this basic simulation, we have also performed two further refinements. Namely, as noted in Ref. [2], twilight events effectively shorten the dead time As discussed above the APD is quenched for 12 ns and then recovers during 5-6 ns which suggests that the effective dead time may be somewhere between 12 and 18 ns. We find a substantially better fit between the model and simulations when we set the dead time to 13.8 ns in our simulation (Table 1, simulation 2).

However, we continue to find that there is disagreement between observations and predictions above 5 MHz detection rate. We find that the origin for this discrepancy is a rate-dependence of the dead time. The apparent dead time is constant and equal to 20 ns up to a detection rate of 5 MHz but becomes longer approximately linear with frequency, reaching 23 ns at a detection rate of 10 MHz. The reason for this rate-dependent dead time is that some capacitances and inductances in the electronic circuit have no time to discharge completely between subsequent detections when the photons arrive too quickly. We accounted for this in our model by changing the dead time in the subroutine linearly with the frequency in the region between 5 and 10 MHz. Results are shown in the last column of Table 1.

Figure 5 shows ratios of simulated values to measured ones for all three simulations. In order to quantify the quality of simulations we calculate the root mean square error of simulated values for *k*-th simulation (*k*=1,2,3) as:

$$R_k = \sqrt{\sum_{i=1}^{N} \frac{(s_{k,i} - m_i)^2}{N}}$$

Where $s_{k,i}$ is *i*-th frequency point in *k*-th simulation and $m_i$ is *i*-th measured correlation. There are *N*=14 points. We obtain: $R_1$=0.0163, $R_2$=0.0052 and $R_3$=0.0009.

Final simulation (*k*=3) achieves the smallest RMS error of only 9E-4, averaged over 4 orders of magnitude of the detection rate.

## 5. Conclusion

Random number generation based on Poissonian distributed stream of photons, a beam splitter and two single-photon detectors suffers from serial autocorrelation errors which do not vanish in either low or high detection frequency limit and moreover change sign at a certain frequency. In this work we have presented an original Monte Carlo simulation of single-photon detectors which successfully reproduces this complex behavior. We believe that this type of simulation can be useful in wide variety of other applications including quantum cryptography, quantum computation and metrology.

## Acknowledgements

We gratefully acknowledge the financial support of the DARPA Defense Sciences Office (DSO) InPho program and Ministry of science education and sports of Republic of Croatia, contract number 098-0352851-2873.

## Notice about authors

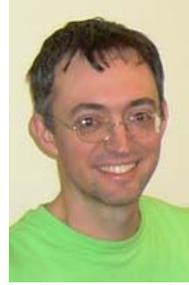

Dr. Mario Stipcevic is a senior scientific associate at the Rudjer Boskovic Institute in Zagreb, Croatia (RBI) and the author of over 48 scientific articles in physics, 17 popular articles in the field of electronics, 3 patent applications and 1 granted patent. He obtained his PhD in high energy physics working on CERN's experiment Atlas in 1994 at L'Universite de Savoie, Chambery, France. Since 2007 he has been leading a research project "Experimental techniques of quantum communication and quantum information" at RBI, URL: http://www.irb.hr/users/stipcevi. His current research interests are: quantum information, quantum cryptography, two-photon entanglement, random number generators and APD based single photon detectors. During 2009/2010 he was a Fulbright scholar at University of California Santa Barbara (UCSB) and in continuation during sabbatical leave he has been working at UCSB and Duke University as research collaborator on projects involving superconducting random numbers and very high speed quantum cryptography. Membership: CERN, Croatian Physical Society and Optical Society of America.

Daniel J. Gauthier is the Robert C. Richardson Professor of Physics at Duke University. He received the B.S., M.S., and Ph.D. degrees from the University of Rochester, Rochester, NY, in 1982, 1983, and 1989, respectively. His Ph.D. research on "Instabilities and chaos of laser beams propagating through nonlinear optical media" was supervised by Prof. R.W. Boyd and supported in part through a University Research Initiative Fellowship. From 1989 to 1991, he developed the first CW two-photon optical laser as a Post-Doctoral Research Associate under the mentorship of Prof. T. W. Mossberg at the University of Oregon. In 1991, he joined the faculty of Duke University, Durham, NC, as an Assistant Professor of Physics and was named a Young Investigator of the U.S. Army Research Office in 1992 and the National Science Foundation in 1993. He was chair of the Duke Physics Department from 2005 - 2011 and is a founding member of the Duke Fitzpatrick Institute for Photonics. His research interests include: high-rate quantum communication, nonlinear quantum optics, single-photon all-optical switching, applications of slow light in classical and quantum information processing, and synchronization and control of the dynamics of complex networks in complex electronic and optical systems. Prof. Gauthier is a Fellow of the Optical Society of America and the American Physical Society.